\documentclass[
 aip, cha,
 amsmath,amssymb,
 reprint,%
]{revtex4-1}
\usepackage{upgreek}
\usepackage{graphicx}
\usepackage{dcolumn}
\usepackage{bm}
\usepackage[utf8]{inputenc}
\usepackage[T1]{fontenc}
\usepackage{mathptmx}
\usepackage{etoolbox}
\usepackage{units}
\usepackage{xcolor}
\definecolor{linkscolor}{RGB}{10,55,130}
\usepackage[pdftex, colorlinks=true, linkcolor=linkscolor, citecolor=linkscolor, urlcolor=linkscolor, breaklinks=true]{hyperref}

\makeatletter
\def\@email#1#2{%
 \endgroup
 \patchcmd{\titleblock@produce}
  {\frontmatter@RRAPformat}
  {\frontmatter@RRAPformat{\produce@RRAP{*#1\href{mailto:#2}{#2}}}\frontmatter@RRAPformat}
  {}{}
}%
\makeatother
\begin{document}

\title{Symmetry-Controlled Ultrastrong Phonon--Photon Coupling in a Terahertz Cavity}

\author{Dasom Kim}
\affiliation{Department of Electrical and Computer Engineering, Rice University, Houston, TX 77005, USA}

\author{Maxime Dherb\'ecourt}
\affiliation{Institut de Physique et Chimie des Mat\'{e}riaux de Strasbourg (UMR 7504), Universit\'e de Strasbourg and CNRS, Strasbourg, 67200, France}

\author{Sae R.\ Endo}
\affiliation{Smalley--Curl Institute, Rice University, Houston, TX 77005, USA}
\affiliation{Department of Applied Physics and Physico-Informatics, Keio University, Yokohama 223-8522, Japan}

\author{Geon Lee}%
\affiliation{Sensor System Research Center, Korea Institute of Science and Technology, Seoul, 02792, Republic of Korea}

\author{Ayush Agrawal}
\affiliation{Department of Chemical and Biomolecular Engineering, Rice University, Houston, TX 77005, USA}

\author{Sunghwan Kim}
\affiliation{Department of Physics, Ulsan National Institute of Science and Technology (UNIST), Ulsan, 44919, Republic of Korea}

\author{Wen-Hua~Wu}
\affiliation{Department of Electrical and Computer Engineering, Rice University, Houston, TX 77005, USA}
\affiliation{Applied Physics Graduate Program, Smalley--Curl Institute, Rice University, Houston, TX 77005, USA}

\author{Aditya~D.~Mohite}
\affiliation{Smalley--Curl Institute, Rice University, Houston, TX 77005, USA}
\affiliation{Department of Chemical and Biomolecular Engineering, Rice University, Houston, TX 77005, USA}
\affiliation{Department of Materials Science and NanoEngineering, Rice University, Houston, TX 77005, USA}

\author{Minah Seo}
\affiliation{Sensor System Research Center, Korea Institute of Science and Technology, Seoul, 02792, Republic of Korea}
\affiliation{Department of Physics, Sogang University, Seoul, 04107, Republic of Korea}

\author{David Hagenm\"uller}
\affiliation{Institut de Physique et Chimie des Mat\'{e}riaux de Strasbourg (UMR 7504), Universit\'e de Strasbourg and CNRS, Strasbourg, 67200, France}

\author{Junichiro Kono*}
 \email{kono@rice.edu}
\affiliation{Department of Electrical and Computer Engineering, Rice University, Houston, TX 77005, USA}
\affiliation{Smalley--Curl Institute, Rice University, Houston, TX 77005, USA}
\affiliation{Department of Materials Science and NanoEngineering, Rice University, Houston, TX 77005, USA}
\affiliation{Rice Advanced Materials Institute, Rice University, Houston, TX 77005, USA}
\affiliation{Department of Physics and Astronomy, Rice University, Houston, TX 77005, USA}

\date{\today}

\begin{abstract}
Optical cavities provide a powerful means to engineer light--matter hybrid states by coupling confined electromagnetic fields with matter excitations. Achieving \textit{in situ} control of the coupling strength is essential for investigating how such hybridization evolves with the coupling strength. In this work, we use a symmetry-changing structural phase transition in lead halide perovskites to reversibly tune the phonon–photon coupling strength, leveraging the fact that their phonon frequencies and oscillator strengths are dictated by lattice symmetry. Terahertz time-domain spectroscopy of MAPbI$_3$ embedded in nanoslot cavities reveals three polariton branches above the critical temperature $T_\mathrm{c} \simeq 162.5$\,K, and the emergence of an additional branch below $T_\mathrm{c}$, activated by a new phonon mode in the low-temperature phase. The full dispersion is accurately reproduced using a multimode Hopfield model, confirming that all normalized coupling strengths remain in the ultrastrong coupling regime. 
These results demonstrate symmetry-controlled tuning of ultrastrong coupling 
via phonon engineering in optical cavities.
\end{abstract}

\maketitle
\section{Introduction}
Light confined in optical cavities can strongly interact with collective excitations in solids, forming mixed light--matter hybrids known as polaritons. The ultrastrong coupling (USC) regime is reached when the light--matter coupling strength $g$ becomes a sizable fraction of the bare excitation frequency~\cite{FriskKockum2019, FornDiaz2019}. USC has now been demonstrated in a variety of solid-state platforms, including Landau polaritons~\cite{Scalari2012,Zhang2016,Li2018NP,Tay2025}, magnetic systems~\cite{Li2018,Makihara2021,Kim2025}, and phonon polaritons~\cite{Kim2020,barra-burilloMicrocavityPhononPolaritons2021,Roh2023,Baydin2025,kim2025_nc}. In this nonperturbative regime, the ground state is predicted to be strongly dressed by vacuum fluctuations, giving rise to associated phenomena such as squeezed ground states~\cite{Ciuti2005,Hayashida2023} and to modifications of vibrational~\cite{wangPhaseTransitionPerovskite2014,Jarc2023}, electronic~\cite{Appugliese2022,thomasExploringSuperconductivityStrong2025}, and magnetic properties~\cite{Thomas2021LargeEnhancementFerromagnetismYBCO,Kritzell2024-rw,Kim2025}. 

A central challenge in rigorously identifying such polaritonic effects is to achieve the ability to vary the coupling strength $g$ \textit{in situ}, so that changes in material properties can be directly linked to changes in $g$. In most cavity--matter systems, however, $g$ is fixed once the cavity geometry and mode volume are defined, limiting experimental control. Attempts to tune $g$ by continuously modifying intrinsic material parameters---such as carrier density~\cite{Paravicini2017} or spin density~\cite{Kritzell2024-rw}---introduce ambiguity, since conventional material responses cannot be cleanly separated from genuine polaritonic effects.

An alternative strategy is to exploit a well-defined phase transition as a built-in tuning parameter~\cite{Jarc2023}. Lead halide perovskites (MAPbX$_3$; MA = CH$_3$NH$_3^+$, X = Cl, Br, I), widely investigated for photovoltaic applications~\cite{Kojima2009}, are particularly well suited in this regard. In MAPbI$_3$, temperature-driven structural transitions between orthorhombic, tetragonal, and cubic phases~\cite{poglitschDynamicDisorderMethylammoniumtrihalogenoplumbates1987,onoda-yamamuroCalorimetricIRSpectroscopic1990,La-o-vorakiat2016} lead to discrete changes in lattice symmetry and phonon spectra, providing a clear reference against which the influence of light--matter hybridizations on phase stability, dynamics, or material properties can be assessed.

Lead halide perovskites have recently emerged as promising systems for realizing USC between photons and optical phonons in terahertz (THz) metamaterial cavities~\cite{Kim2020,Roh2023,kim2025_nc}. While previous studies have demonstrated USC within a single structural phase, and one recent work has investigated phonon--photon coupling across a phase transition in the strong-coupling regime~\cite{Jarc2024}, the evolution of polariton formation under USC across a symmetry-changing transition remains largely unexplored. 

In this work, we use a structural phase transition in MAPbI$_3$ thin films as a reversible control knob to modulate $g$ in subwavelength-mode-volume THz nanoslot cavities. Temperature-dependent THz time-domain spectroscopy (THz-TDS) reveals the emergence of an additional polariton branch below the critical temperature $T_\mathrm{c} \simeq 162.5\,\mathrm{K}$, consistent with the activation of a new phonon mode in the orthorhombic phase. The full polariton dispersion across the transition is accurately reproduced using a multimode Hopfield model, confirming that all normalized coupling strengths remain in the USC regime. Within our experimental precision, the transition temperature is unchanged, indicating that light--matter hybridization does not measurably affect phase stability under the present coupling conditions. These results provide new insight into multimode polariton formation across phase boundaries and establish a pathway for engineering functional material properties in perovskites and related systems via ultrastrong light--matter interactions.

\section{Results}

At room temperature, MAPbI$_3$ adopts a tetragonal structure and exhibits two phonon modes in the THz range, originating from the rocking and stretching vibrations of the Pb--I bonds~\cite{poglitschDynamicDisorderMethylammoniumtrihalogenoplumbates1987, onoda-yamamuroCalorimetricIRSpectroscopic1990, La-o-vorakiat2016}. Upon cooling, the material undergoes a structural phase transition at $T_\text{c} \approx 162.5\,\text{K}$ from tetragonal to orthorhombic symmetry. This symmetry change modifies the vibrational spectrum of the Pb--I framework, causing the two tetragonal-phase phonon modes to split into four distinct modes in the orthorhombic phase~\cite{La-o-vorakiat2016}.

\begin{figure}[t!]
\centering
\includegraphics[width=8cm]{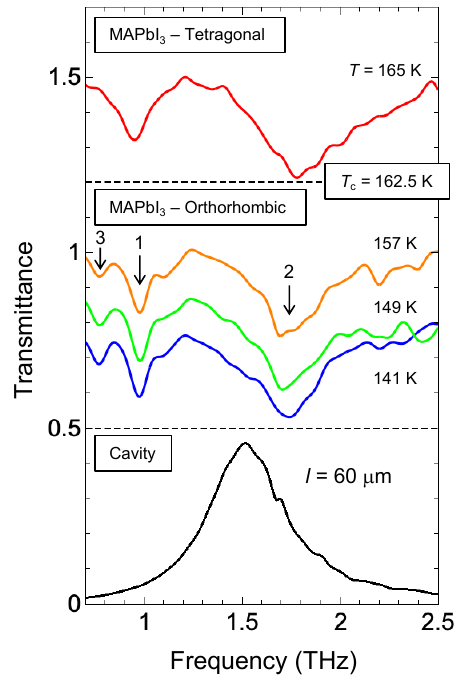} 
\caption{Temperature-dependent terahertz transmission spectra of MAPbI$_3$. Upper: Bare MAPbI$_3$ film showing two phonon modes at around 0.95\,THz and 1.78\,THz in the tetragonal phase (165\,K) and three modes in the orthorhombic phase (below $T_\text{c}$ = 162.5\,K) due to structural splitting of Pb--I vibrations. Lower: Transmission of a nanoslot cavity ($l =$ 60\,$\upmu$m) exhibiting a resonance at 1.52\,THz.} 
\label{fig1}
\end{figure}

To verify this behavior, we performed THz-TDS on a 200\,nm-thick bare MAPbI$_3$ film. Figure~\ref{fig1} (upper panel) shows the THz transmission spectra in the two structural phases. In the tetragonal phase (165\,K), two absorption dips are observed at approximately 0.95\,THz and 1.78\,THz, corresponding to the TO$_1$ and TO$_2$ phonon modes. Below $T_\text{c}$, the lower mode splits into two peaks at 0.77\,THz and 0.98\,THz, while the TO$_2$ mode appears unchanged. The expected splitting of the TO$_2$ mode is not resolved in the spectra because the oscillator strength of the fourth mode is significantly weaker than that of the original TO$_2$ mode in this temperature range~\cite{La-o-vorakiat2016}.

\begin{figure*}[t!]
\centering
\includegraphics[width=16cm]{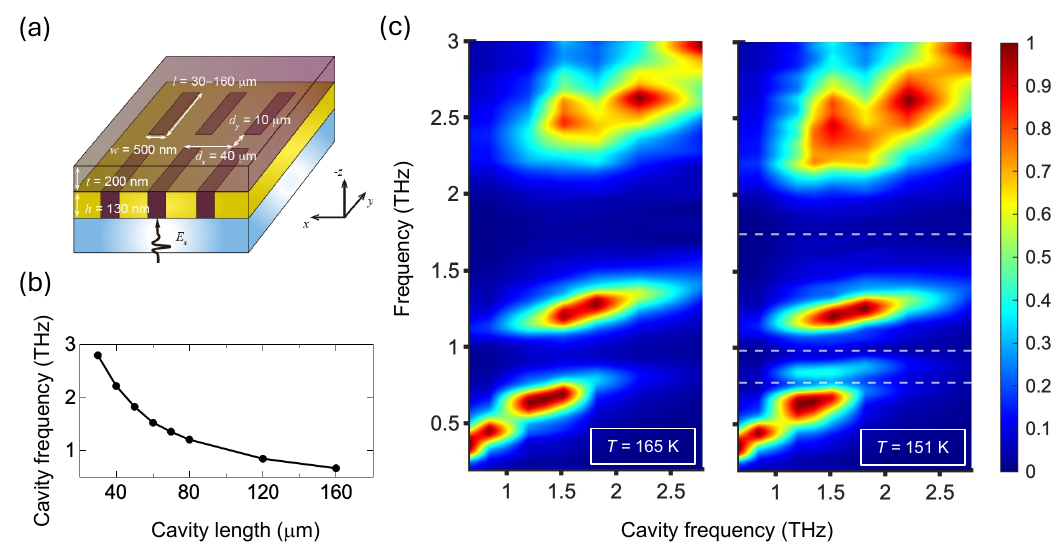} 
\caption{THz transmission maps of MAPbI$_3$--nanoslot hybrid systems. (a)~Schematic illustration of the hybrid structure. (b)~The dependence of the cavity resonance frequency on nanoslot length $l$. (c)~Normalized transmittance color maps at 165\,K (tetragonal) and 151\,K (orthorhombic), respectively. Three polaritonic branches, lower (LP), middle (MP), and upper (UP), arise from the ultrastrong coupling of the cavity mode with phonon modes TO$_1$ and TO$_2$; below $T_\text{c}$ a new branch appears near 0.83\,THz, indicating coupling with the emergent third phonon mode.} 
\label{Fig2}
\end{figure*}

Because the MAPbI$_3$ film thickness is much smaller than the THz wavelength, the bare THz electromagnetic field couples only weakly to the phonons. To reach the USC regime, we employ nanoslot cavities that strongly enhance the electric field inside the slots~\cite{Seo2009,Kim2018}. The cavities were fabricated as arrays of nanoslots with a width of $w = 500\,\text{nm}$ on quartz substrates using standard photolithography and lift-off techniques; see Fig.\,\ref{Fig2} for fabrication details. Figure~\ref{fig1} (lower panel) shows a representative transmittance spectrum for a nanoslot with length $l = 60\,\upmu\text{m}$, featuring a single peak at the resonance frequency.

Next, we deposited 200-nm-thick MAPbI$_3$ films onto the nanoslot arrays. A schematic of the resulting MAPbI$_3$--nanoslot hybrid structures is shown in Fig.\,\ref{Fig2}(a). To tune the cavity resonance, we fabricated samples with different cavity lengths ($l = 30, 40, 50, 60, 70, 80, 120, 160\,\upmu\text{m}$). The dependence of the cavity frequency on $l$ is plotted in Fig.\,\ref{Fig2}(b). Because the resonance is primarily determined by the nanoslot geometry, the cavity frequency decreases as $l$ increases. By varying $l$, we therefore control the detuning between the bare cavity mode and the phonon mode, and consequently the degree of light--matter hybridization for each phonon branch.

We carried out THz-TDS measurements on the hybrid systems at various temperatures across $T_\text{c}$. Time-domain signals from the samples and a quartz reference were recorded and Fourier transformed to obtain transmission spectra in the range 0.2--3\,THz. Figure~\ref{Fig2}(c) presents color maps of the normalized transmittance for hybrid systems with different cavity frequencies $\omega_\text{c}$ (set by $l$) in the tetragonal phase (left, 165\,K) and orthorhombic phase (right, 151\,K). In the tetragonal phase (165\,K), three polariton branches are observed, consistent with previous results~\cite{kim2025_nc}. These features correspond to the lower- (LP), middle- (MP), and upper-polariton (UP) modes arising from the hybridization of the TO$_1$ and TO$_2$ phonons with the cavity mode, as shown below. The larger oscillator strength of TO$_2$ leads to a correspondingly larger Rabi splitting. In the orthorhombic phase (151\,K), an additional branch appears near 0.83\,THz for $1.2\,\text{THz} \lesssim \omega_\text{c} \lesssim 1.8\,\text{THz}$. The three branches below 1.5\,THz are now separated by the two lower-frequency phonon modes (labeled 1 and 3 in Fig.\,\ref{fig1}), as indicated by the lower dashed white lines. We show below that the additional polariton branch arises from the coupling between the cavity mode and the third phonon mode, which becomes active in the orthorhombic phase, at low cavity frequencies, and with the first phonon mode at high cavity frequencies. In addition, the UP branch becomes broader and red-shifted, a behavior attributed to the emergence of the fourth phonon mode.

The formation of phonon-polaritons can be described theoretically by a multimode Hopfield model\cite{hopfield_theory_1958}. Following the approach of Ref.~[\onlinecite{kim2025_nc}], we start from the microscopic Hamiltonian
\begin{eqnarray}\label{eq:hamiltonian}
    H &=& \hbar \omega_{\mathrm{c}}\, a^\dagger a + \sum_\lambda \hbar \omega_\lambda\, b_\lambda^\dagger b_\lambda - \mathrm{i} \sum_\lambda \hbar g_\lambda (b_\lambda - b_\lambda^\dagger)(a + a^\dagger) \nonumber \\
    &&+ \sum_{\lambda} \frac{\hbar g_\lambda^2}{\omega_\lambda}\,(a + a^\dagger)^2,
\end{eqnarray}
where $a$ ($a^\dagger$) is the annihilation (creation) operator of the cavity photon mode with frequency $\omega_{\mathrm{c}}$, and $b_\lambda$ ($b_\lambda^\dagger$) the annihilation (creation) operator of the phonon mode labeled by $\lambda$ and with frequency $\omega_\lambda$.
For temperatures above $T_\text{c}$ we include two phonon modes, $\lambda=1,2$, corresponding to the $\textrm{TO}_1$ and $\textrm{TO}_2$ transverse optical phonons of the perovskite, while below $T_\text{c}$ an additional mode $\lambda=3$ is incorporated to account for the phonon that appears in the orthorhombic phase.
The first two terms in Eq.~\eqref{eq:hamiltonian} describe the free energy of the cavity field and of the phonon modes, respectively. The third term represents the light--matter interaction, with coupling strength $\lambda$ given by $g_\lambda = \nu_\lambda \sqrt{\omega_\lambda/\omega_\mathrm{c}} / 2$. Here, $\nu_\lambda$ corresponds to the effective ionic plasma frequency of mode $\lambda$, characterizing the effective charges associated with the $\mathrm{Pb}^{2+}$ and $\mathrm{I}^{-}$ ions participating in the lattice vibration. The final term is the diamagnetic (or $A^2$) contribution, which becomes important in the USC regime (see Ref.~[\onlinecite{kim2025_nc}] for derivation details).

Equation~\eqref{eq:hamiltonian} can be diagonalized by performing a Hopfield-Bogoliubov transformation and introducing bosonic polaritonic annihilation operators of the form $p_\alpha = w_\alpha\, a + \sum_\lambda x_{\lambda,\alpha}\, b_\lambda + y_\alpha\, a^\dagger + \sum_\lambda z_{\lambda,\alpha}\, b_\lambda^\dagger$, where $w_\alpha$, $x_{\lambda,\alpha}$, $y_\alpha$, and $z_{\lambda,\alpha}$ are complex coefficients. These operators allow the Hamiltonian to be expressed as $H = \sum_\alpha \hbar \Omega_\alpha\, p_\alpha^\dagger p_\alpha$, with $\Omega_\alpha$ denoting the polaritonic eigenfrequencies.
The index $\alpha$ runs over the total number of polaritonic modes, i.e., the single cavity degree of freedom together with the phonon modes present in the phase under consideration.
In the USC regime the diamagnetic ($A^2$) contribution in Eq.~\eqref{eq:hamiltonian} becomes significant and can no longer be neglected, as well as the anti‑resonant components of the Hopfield transformation, contained in the coefficients $y_\alpha$ and $z_{\lambda,\alpha}$.
Accordingly, each polaritonic mode $\alpha$ contains contributions from each individual phonon $\textrm{TO}_\lambda$ quantified by the phonon fraction $\mathcal{F}^\mathrm{ph}_{\lambda,\alpha} = |x_{\lambda,\alpha}|^2 - |z_{\lambda,\alpha}|^2$, and from the cavity photons through the fraction $\mathcal{F}^\mathrm{pt}_{\alpha} = |w_\alpha|^2 - |y_\alpha|^2$. Here, both phononic and photonic weights explicitly account for the contributions of both the resonant and anti‑resonant Hopfield coefficients.

\begin{figure}
    \centering
    \includegraphics[width=\columnwidth]{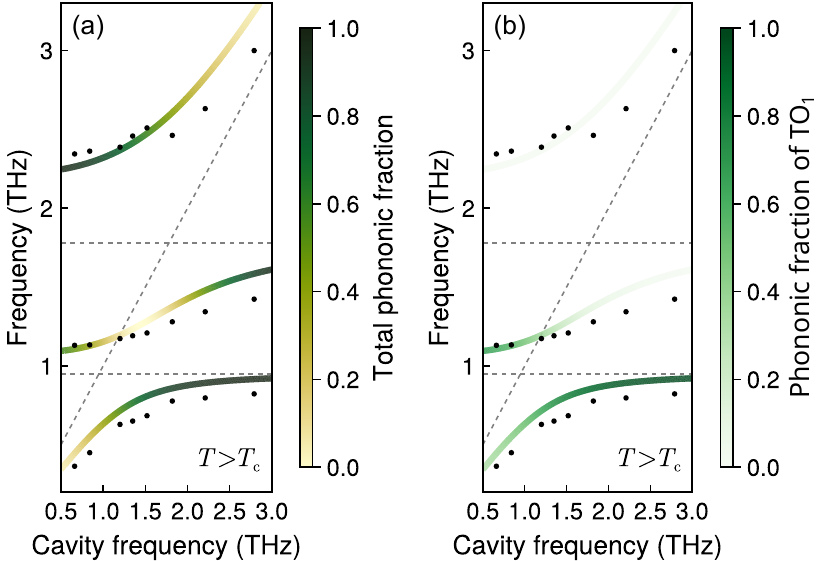}
    \caption{\label{fig:fitting_data_hopfield_high_t} Polaritonic dispersion as a function of the cavity frequency for $T =\unit[165]{K} > T_\textrm{c}$, calculated by numerical diagonalization of Eq.~\eqref{eq:hamiltonian}, and compared with the experimental data (black markers).
    (a) The colour scale encodes the photonic (yellow) and phononic (green) content of each polaritonic eigenmode $\alpha$, quantified by the fractions $\sum_\lambda \mathcal{F}^\mathrm{ph}_{\lambda,\alpha}$ and $\mathcal{F}^\mathrm{pt}_{\alpha}$, respectively.
    In (b), the colours scale represents the phononic contribution from the $\textrm{TO}_1$ mode to each polariton branch $\mathcal{F}^\mathrm{ph}_{\lambda=1,\alpha}$.
    (a, b) Dashed lines indicate the uncoupled cavity and bare phonon-mode frequencies.
    The transverse--optical phonons are located at $\omega_1=\unit[0.95]{THz}$ and $\omega_2=\unit[1.78]{THz}$.
    The coupling strengths $g_\lambda$ reported in the main text are then obtained by fitting the calculated dispersion to these experimental data.}
\end{figure}

\begin{figure}[t!]
    \centering
    \includegraphics[width=\columnwidth]{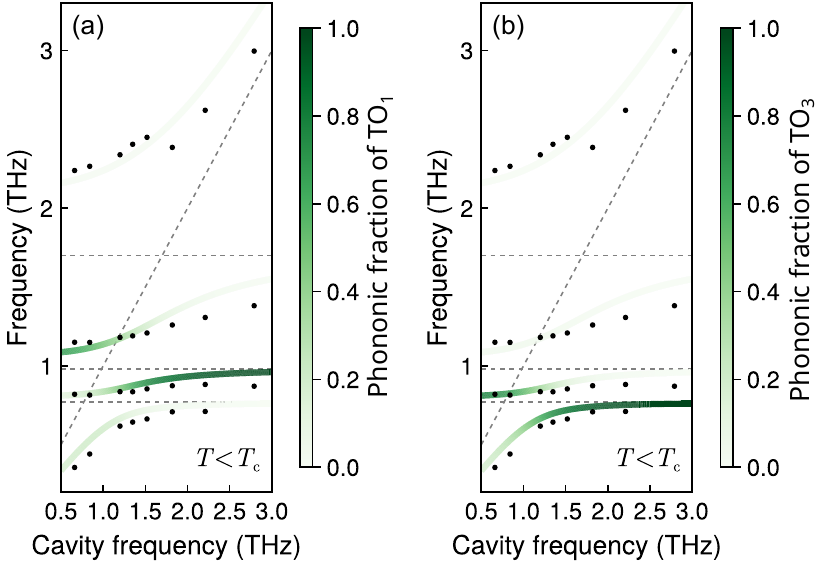}
    \caption{\label{fig:fitting_data_hopfield_low_t} Polaritonic dispersion as a function of the cavity frequency, calculated by numerical diagonalization of Eq.~\eqref{eq:hamiltonian}, and compared with the experimental data (black markers), simirally as Fig.~\ref{fig:fitting_data_hopfield_high_t}, but for $T =\unit[151]{K} < T_\textrm{c}$.
    The colours scale represents the phononic contribution from the $\textrm{TO}_1$ (a) and $\textrm{TO}_3$ (b) modes to each polariton branch.
    Here, the transverse--optical phonons are located at $\omega_1=\unit[0.98]{THz}$, $\omega_2=\unit[1.7]{THz}$ and $\omega_3=\unit[0.77]{THz}$.}
\end{figure}

Numerical diagonalization of the Hamiltonian was performed for a range of cavity frequencies $\omega_{\mathrm{c}}$, and the resulting polaritonic dispersions were fitted to the experimental data by varying the light--matter coupling strengths $g_\lambda$, while using the experimentally determined phonon frequencies $\omega_\lambda$ as fixed inputs (see Figs.~\ref{fig:fitting_data_hopfield_high_t} and \ref{fig:fitting_data_hopfield_low_t}).
Each coupling strength $g_\lambda$ is measured numerically at resonance between the cavity and the corresponding TO frequency, that is $\omega_{\mathrm{c}}=\omega_\lambda$, for every phonon mode $\lambda$.
From this procedure, we obtain the following normalized couplings. Above the transition temperature $T_\text{c}$ (tetragonal phase) the $\textrm{TO}_1$ and $\textrm{TO}_2$ modes yield $g_1/\omega_1=0.36$ and $g_2/\omega_2=0.35$.
Below $T_\text{c}$ (orthorhombic phase), where an additional transverse--optical mode $\textrm{TO}_3$ is present, the extracted values are $g_1 / \omega_1=0.28$, $g_2 / \omega_2=0.36$, and $g_3 / \omega_3=0.25$.
All reported couplings lie within the USC regime with the nanoslot resonator, which underlines the necessity of retaining the diamagnetic ($A^2$) term in Eq.~\eqref{eq:hamiltonian} as well as the anti‑resonant terms.
The temperature dependence of the extracted couplings (Figs.~\ref{fig:fitting_data_hopfield_high_t} and \ref{fig:fitting_data_hopfield_low_t}) reveals a marked variation at the structural phase transition.
While the coupling of the $\textrm{TO}_2$ mode is essentially temperature independent, the coupling of the $\textrm{TO}_1$ mode decreases by approximately 30\% upon cooling from the tetragonal to the orthorhombic phase. The emergent $\textrm{TO}_3$ mode below $T_\text{c}$ exhibits a coupling strength comparable to that of $\textrm{TO}_1$. These changes are attributed to the lattice reorganization across the phase transition, which modifies the effective ionic plasma frequencies $\nu_\lambda$ entering the light--matter interaction, while preserving the system in the USC regime.
Now examining the phononic and photonic weights, as shown in Fig.~\ref{fig:fitting_data_hopfield_high_t} above the critical temperature, we find that the $\textrm{TO}_1$ mode predominantly contributes to the phononic character of the two lowest polaritonic branches. Below $T_\mathrm{c}$ (Fig.~\ref{fig:fitting_data_hopfield_low_t}), the emergence of the new $\textrm{TO}_3$ mode leads to a redistribution of the total phononic weight between the LP and the newly formed MP branch just below $1$\,THz. While this branch is primarily composed of the $\textrm{TO}_1$ mode at high cavity frequencies, it mainly consists of the $\textrm{TO}_3$ mode at low cavity frequencies.

To examine whether the formation of light--matter hybrids in the USC regime influences the structural phase transition in MAPbI$_3$, we analyzed the evolution of polariton frequencies as a function of temperature for two different cavity frequencies $\omega_\text{c} = 1.2$\,THz and 2.2\,THz. We note that the bare film response is too weak to resolve such detailed temperature evolution under our experimental conditions, making the cavity-coupled system essential for sensitively tracking the transition through polaritonic features.
\begin{figure}[t!]
\centering
\includegraphics[width=8.5cm]{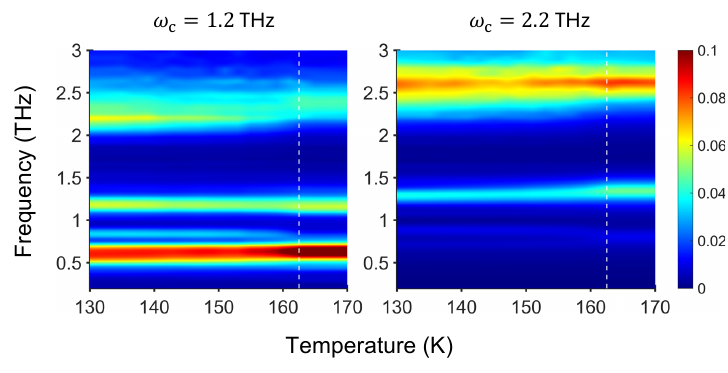} 
\caption{Temperature-dependent THz transmission spectra of hybrid systems with cavity frequencies $\omega_\text{c}=1.2$ THz and 2.2\,THz. Each color map shows the evolution of polariton branches as the sample is heated through the tetragonal–orthorhombic transition near $T_\text{c}\simeq$ 162.5\,K, indicated by vertical dashed lines. Some of the polariton branches gradually disappear or saturate above $T_\text{c}$. Despite the markedly different light--matter hybridization conditions, the transition occurs at approximately the same temperature within the 2\,K experimental uncertainty.} 
\label{fig4}
\end{figure}
In Fig.~\ref{fig:fitting_data_hopfield_high_t}(a) ($T > T_\mathrm{c}$), the photonic weight of each polariton branch, indicated by the color scale, reveals that for $\omega_\text{c} = 1.2$\,THz, the two lower polaritonic branches possess the highest photonic fraction, whereas at $\omega_\text{c} = 2.2$\,THz, the upper polariton (UP) branch carries most of the photonic weight.
Below the critical temperature [$T < T_\mathrm{c}$, see Fig.~\ref{fig:fitting_data_hopfield_low_t}], these relative trends are preserved. However, the emergence of two new polariton modes below $1$\,THz leads to a partial transfer of phononic weight from the original LP branch, thereby reducing their individual phonon contributions.
As shown in Fig.\,\ref{fig4}, the phase transition from the orthorhombic to tetragonal phase is clearly tracked by the spectral evolution and the disappearance of polariton branches across $T_\text{c}$ $\simeq$ 162.5\,K, indicated by the vertical dashed lines. In both cases, the UP branches show the saturation behavior at above 162.5\,K. Similarly, other polariton branches show the disappearance or saturation across $T_\text{c}$. Despite the markedly different hybridization, the disappearance of the orthorhombic-phase polariton features occurred at approximately the same $T_\text{c}$ within our experimental resolution of 2\,K. These results suggest that, within our experimental limit, the formation of phonon-polaritons in the USC regime does not noticeably alter the structural phase transition temperature in MAPbI$_3$. Nonetheless, the possibility of more subtle cavity-induced effects, such as modified lattice fluctuations, cannot be excluded and may warrant further investigation under different coupling geometries or stronger vacuum-field confinement. In particular, coupling to a soft phonon mode that approaches zero frequency at the phase boundary could provide an intriguing route to explore cavity-modified critical behavior. 

\section{Conclusion}
In summary, we have demonstrated symmetry-controlled multimode USC between THz cavity photons and optical phonons in hybrid MAPbI$_3$--nanoslot systems. Above $T_\text{c}$, the tetragonal phase exhibits two infrared-active phonons that hybridize with the cavity mode to form three distinct polariton branches. Below $T_\text{c}$, the transition to the orthorhombic phase introduces an additional phonon mode, leading to the emergence of a new polariton branch and a redistribution of the coupling strength. Overall, this work establishes halide perovskites as a versatile platform for exploring light--matter interactions with the \textit{in-situ} control of the coupling strength $g$, paving the way toward phonon-engineered quantum optoelectronic materials.

\begin{acknowledgments}
D.K. and J.K.\ acknowledge support from the U.S.\ Army Research Office (through Award No.\ W911NF2110157), the W.\ M.\ Keck Foundation (through Award No.\ 995764), the Gordon and Betty Moore Foundation (through Grant No.\ 11520), and the Robert A.\ Welch Foundation (through Grant No.\ C-1509).
M.D. and D.H.\ acknowledge support from the Interdisciplinary Thematic Institute QMat, as part of the ITI 2021-2028 program of the University of Strasbourg, CNRS, and Inserm, supported by IdEx Unistra (Project No.~ANR 10 IDEX 0002), and by SFRI STRAT’US Projects No.~ANR-20-SFRI-0012 and No.~ANR-17-EURE-0024 under the framework of the French Investments for the Future Program.
\end{acknowledgments}

\section*{Author contributions}
J.K.\ supervised the project. D.K.\ conceived the project, built the THz spectroscopy setup, performed THz measurements, and analyzed the experimental data under the guidance of J.K. S.R.E.\ and W.-H.W.\ helped THz measurements. G.L., S.K., and D.K.\ designed and fabricated the nanoslot cavities under the guidance of M.S. M.D.\ and D.H.\ developed the theoretical model and performed the calculations. A.A.\ grew the perovskite films under the guidance of A.D.M. D.K., M.D., S.R.E., D.H., and J.K.\ wrote the manuscript. All authors discussed the results and commented on the manuscript.

\section*{Data Availability Statement}
Data available on request from the authors.

\bibliography{results, ref}

@article{Thomas2021LargeEnhancementFerromagnetismYBCO,
  author    = {Thomas, Anoop and Devaux, {'E}lo{"i}se and Nagarajan, Kalaivanan and Rogez, Guillaume and Seidel, Marcus and Richard, Fanny and Genet, Cyriaque and Drillon, Marc and Ebbesen, Thomas W.},
  title     = {Large Enhancement of Ferromagnetism under a Collective Strong Coupling of {YBa$_2$Cu$_3$O$_{7-x}$} Nanoparticles},
  journal   = {Nano Letters},
  year      = {2021},
  volume    = {21},
  number    = {8},
  pages     = {4365--4370},
  doi       = {10.1021/acs.nanolett.1c00973}
}

@article{barra-burilloMicrocavityPhononPolaritons2021,
  title = {Microcavity Phonon Polaritons from the Weak to the Ultrastrong Phonon--Photon Coupling Regime},
  author = {{Barra-Burillo}, Mar{\'i}a and Muniain, Unai and Catalano, Sara and Autore, Marta and Casanova, F{\`e}lix and Hueso, Luis E. and Aizpurua, Javier and Esteban, Ruben and Hillenbrand, Rainer},
  year = 2021,
  month = oct,
  journal = {Nature Communications},
  volume = {12},
  number = {1},
  pages = {6206},
  issn = {2041-1723},
  doi = {10.1038/s41467-021-26060-x},
  url = {https://www.nature.com/articles/s41467-021-26060-x}
}

@article{onoda-yamamuroCalorimetricIRSpectroscopic1990,
  title = {Calorimetric and {{IR}} Spectroscopic Studies of Phase Transitions in Methylammonium Trihalogenoplumbates ({{II}})\dag},
  author = {{Onoda-Yamamuro}, Noriko and Matsuo, Takasuke and Suga, Hiroshi},
  year = 1990,
  month = jan,
  journal = {Journal of Physics and Chemistry of Solids},
  volume = {51},
  number = {12},
  pages = {1383--1395},
  issn = {00223697},
  doi = {10.1016/0022-3697(90)90021-7},
  url = {https://linkinghub.elsevier.com/retrieve/pii/0022369790900217}
}

@article{poglitschDynamicDisorderMethylammoniumtrihalogenoplumbates1987,
  title = {Dynamic Disorder in Methylammoniumtrihalogenoplumbates ({{II}}) Observed by Millimeter-Wave Spectroscopy},
  author = {Poglitsch, A. and Weber, D.},
  year = 1987,
  month = dec,
  journal = {The Journal of Chemical Physics},
  volume = {87},
  number = {11},
  pages = {6373--6378},
  issn = {0021-9606, 1089-7690},
  doi = {10.1063/1.453467},
  url = {https://pubs.aip.org/jcp/article/87/11/6373/220079/Dynamic-disorder-in}
}

@article{thomasExploringSuperconductivityStrong2025,
  title = {Exploring Superconductivity under Strong Coupling with the Vacuum Electromagnetic Field},
  author = {Thomas, A. and Devaux, E. and Nagarajan, K. and Chervy, T. and Seidel, M. and Rogez, G. and Robert, J. and Drillon, M. and Ruan, T. T. and Schlittenhardt, S. and Ruben, M. and Hagenm{\"u}ller, D. and Sch{\"u}tz, S. and Schachenmayer, J. and Genet, C. and Pupillo, G. and Ebbesen, T. W.},
  year = 2025,
  month = apr,
  journal = {The Journal of Chemical Physics},
  volume = {162},
  number = {13},
  pages = {134701},
  issn = {0021-9606},
  doi = {10.1063/5.0231202},
  url = {https://doi.org/10.1063/5.0231202}
}

@article{wangPhaseTransitionPerovskite2014,
  title = {Phase Transition of a Perovskite Strongly Coupled to the Vacuum Field},
  author = {Wang, Shaojun and Mika, Arkadiusz and Hutchison, James A. and Genet, Cyriaque and Jouaiti, Abdelaziz and Hosseini, Mir Wais and Ebbesen, Thomas W.},
  year = 2014,
  journal = {Nanoscale},
  volume = {6},
  number = {13},
  pages = {7243--7248},
  issn = {2040-3364, 2040-3372},
  doi = {10.1039/C4NR01971G},
  url = {https://xlink.rsc.org/?DOI=C4NR01971G}
}

@Article{Zhang2016,
  author    = {Zhang, Qi and Lou, Minhan and Li, Xinwei and Reno, John L. and Pan, Wei and Watson, John D. and Manfra, Michael J. and Kono, Junichiro},
  title     = {Collective non-perturbative coupling of 2{D} electrons with high-quality-factor terahertz cavity photons},
  journal   = {Nat. Phys.},
  year      = {2016},
  volume    = {12},
  number    = {11},
  pages     = {1005--1011},
  issn      = {1745-2473},
  publisher = {Nature Publishing Group},
}

@Article{FornDiaz2019,
  author    = {Forn-D\'{\i}az, P. and Lamata, L. and Rico, E. and Kono, J. and Solano, E.},
  journal   = {Rev. Mod. Phys.},
  title     = {Ultrastrong coupling regimes of light--matter interaction},
  year      = {2019},
  pages     = {025005},
  volume    = {91},
  doi       = {10.1103/RevModPhys.91.025005},
  issue     = {2},
  numpages  = {48},
  publisher = {American Physical Society},
  url       = {https://link.aps.org/doi/10.1103/RevModPhys.91.025005},
}

@Article{FriskKockum2019,
  author   = {Frisk Kockum, Anton and Miranowicz, Adam and De Liberato, Simone and Savasta, Salvatore and Nori, Franco},
  journal  = {Nat. Rev. Phys.},
  title    = {Ultrastrong coupling between light and matter},
  year     = {2019},
  issn     = {2522-5820},
  number   = {1},
  pages    = {19--40},
  volume   = {1},
  refid    = {Frisk Kockum2019},
  url      = {https://doi.org/10.1038/s42254-018-0006-2},
}

@Article{Hayashida2023,
  author  = {Hayashida, Kenji and Makihara, Takuma and Peraca, Nicolas Marquez and Padilla, Diego Fallas and Pu, Han and Kono, Junichiro and Bamba, Motoaki},
  journal = {Sci. Rep.},
  title   = {Perfect intrinsic squeezing at the superradiant phase transition critical point},
  year    = {2023},
  pages = {2526},
  volume  = {13},
}

@article{Li2018NP,
	author = {Li, Xinwei and Bamba, Motoaki and Zhang, Qi and Fallahi, Saeed and Gardner, Geoff C. and Gao, Weilu and Lou, Minhan and Yoshioka, Katsumasa and Manfra, Michael J. and Kono, Junichiro},
	journal = {Nat. Photon.},
	pages = {324--329},
	title = {Vacuum {B}loch--{S}iegert shift in {L}andau polaritons with ultra-high cooperativity},
	volume = {12},
	year = {2018}}

@article{Makihara2021,
	author = {Makihara, Takuma and Hayashida, Kenji and Noe II, G. Timothy and Li, Xinwei and Marquez Peraca, Nicolas and Ma, Xiaoxuan and Jin, Zuanming and Ren, Wei and Ma, Guohong and Katayama, Ikufumi and Takeda, Jun and Nojiri, Hiroyuki and Turchinovich, Dmitry and Cao, Shixun and Bamba, Motoaki and Kono, Junichiro},
	da = {2021/05/25},
	doi = {10.1038/s41467-021-23159-z},
	id = {Makihara2021},
	journal = {Nat. Commun.},
	number = {1},
	pages = {3115},
	title = {Ultrastrong magnon--magnon coupling dominated by antiresonant interactions},
	ty = {JOUR},
	url = {https://doi.org/10.1038/s41467-021-23159-z},
	volume = {12},
	year = {2021},
	}

@article{La-o-vorakiat2016,
   author = {{La-o-vorakiat}, Chan and Xia, Huanxin and Kadro, Jeannette and Salim, Teddy and Zhao, Daming and Ahmed, Towfiq and Lam, Yeng Ming and Zhu, Jian-Xin and Marcus, Rudolph A. and Michel-Beyerle, Maria-Elisabeth and Chia, Elbert E. M.},
   title = {Phonon Mode Transformation Across the Orthohombic–Tetragonal Phase Transition in a Lead Iodide Perovskite {CH\(_3\)NH\(_3\)PbI\(_3\)}: A Terahertz Time-Domain Spectroscopy Approach},
   journal = {J. Phys. Chem. Lett.},
   volume = {7},
   number = {1},
   pages = {1-6},
   DOI = {10.1021/acs.jpclett.5b02223},
   url = {https://doi.org/10.1021/acs.jpclett.5b02223
https://pubs.acs.org/doi/pdf/10.1021/acs.jpclett.5b02223},
   year = {2016},
   type = {Journal Article}
}

@article{Kim2020,
   author = {Kim, Hwan Sik and Ha, Na Young and Park, Ji-Yong and Lee, Soonil and Kim, Dai-Sik and Ahn, Yeong Hwan},
   title = {Phonon-Polaritons in Lead Halide Perovskite Film Hybridized with {THz} Metamaterials},
   journal = {Nano Lett.},
   volume = {20},
   number = {9},
   pages = {6690-6696},
   ISSN = {1530-6984},
   DOI = {10.1021/acs.nanolett.0c02572},
   url = {https://doi.org/10.1021/acs.nanolett.0c02572
https://pubs.acs.org/doi/pdf/10.1021/acs.nanolett.0c02572},
   year = {2020},
   type = {Journal Article}
}

@article{Roh2023,
   author = {Roh, Yeeun and Chae, Minjun and Oh, Jaewon and Kim, Woochul and Ryu, Mee-Yi and Seo, Minah and Jeong, Jeeyoon},
   title = {Ultrastrong Coupling Enhancement with Squeezed Mode Volume in Terahertz Nanoslots},
   journal = {Nano Lett.},
   ISSN = {1530-6984},
   DOI = {10.1021/acs.nanolett.3c01913},
   url = {https://doi.org/10.1021/acs.nanolett.3c01913},
   volume = {23},
   pages = {7086-7091},
   year = {2023},
   type = {Journal Article}
}

@article{Ciuti2005,
   author = {Ciuti, Cristiano and Bastard, Gérald and Carusotto, Iacopo},
   title = {Quantum vacuum properties of the intersubband cavity polariton field},
   journal = {Phys. Rev. B},
   volume = {72},
   number = {11},
   pages = {115303},
   DOI = {10.1103/PhysRevB.72.115303},
   url = {https://link.aps.org/doi/10.1103/PhysRevB.72.115303
https://journals.aps.org/prb/pdf/10.1103/PhysRevB.72.115303},
   year = {2005},
   type = {Journal Article}
}

@article{Appugliese2022,
   author = {Appugliese, Felice and Enkner, Josefine and Paravicini-Bagliani, Gian Lorenzo and Beck, Mattias and Reichl, Christian and Wegscheider, Werner and Scalari, Giacomo and Ciuti, Cristiano and Faist, Jérôme},
   title = {Breakdown of topological protection by cavity vacuum fields in the integer quantum {H}all effect},
   journal = {Science},
   volume = {375},
   number = {6584},
   pages = {1030-1034},
   DOI = {doi:10.1126/science.abl5818},
   url = {https://www.science.org/doi/abs/10.1126/science.abl5818},
   year = {2022},
   type = {Journal Article}
}

@article{Baydin2025,
      title={Terahertz cavity phonon polaritons in lead telluride in the deep-strong coupling regime}, 
      author={Andrey Baydin and Manukumara Manjappa and Sobhan Subhra Mishra and Hongjing Xu and Jacques Doumani and Fuyang Tay and Dasom Kim and Paulo H. O. Rappl and Eduardo Abramof and Ranjan Singh and Felix G. G. Hernandez and Junichiro Kono},
      year={2025},
      eprint={2501.10856},
      journal = {arXiv:2501.10856},
      primaryClass={physics.optics},
      url={https://arxiv.org/abs/2501.10856}, 
}

@article{Tay2025,
   author = {Tay, Fuyang and Mojibpour, Ali and Sanders, Stephen and Liang, Shuang and Xu, Hongjing and Gardner, Geoff C. and Baydin, Andrey and Manfra, Michael J. and Alabastri, Alessandro and Hagenmüller, David and Kono, Junichiro},
   title = {Multimode ultrastrong coupling in three-dimensional photonic-crystal cavities},
   journal = {Nat. Commun.},
   volume = {16},
   number = {1},
   pages = {3603},
   ISSN = {2041-1723},
   DOI = {10.1038/s41467-025-58835-x},
   url = {https://doi.org/10.1038/s41467-025-58835-x
https://www.nature.com/articles/s41467-025-58835-x.pdf},
   year = {2025},
   type = {Journal Article}
}

@article{Seo2009,
   author = {Seo, M. A. and Park, H. R. and Koo, S. M. and Park, D. J. and Kang, J. H. and Suwal, O. K. and Choi, S. S. and Planken, P. C. M. and Park, G. S. and Park, N. K. and Park, Q. H. and Kim, Dai-Sik},
   title = {Terahertz field enhancement by a metallic nano slit operating beyond the skin-depth limit},
   journal = {Nat. Photon.},
   volume = {3},
   number = {3},
   pages = {152-156},
   ISSN = {1749-4893},
   DOI = {10.1038/nphoton.2009.22},
   url = {https://doi.org/10.1038/nphoton.2009.22},
   year = {2009},
   type = {Journal Article}
}

@article{Kim2018,
   author = {Kim, Dasom and Jeong, Jeeyoon and Choi, Geunchang and Bahk, Young-Mi and Kang, Taehee and Lee, Dukhyung and Thusa, Bidhek and Kim, Dai-Sik},
   title = {Giant Field Enhancements in Ultrathin Nanoslots above 1 Terahertz},
   journal = {ACS Photonics},
   volume = {5},
   number = {5},
   pages = {1885-1890},
   DOI = {10.1021/acsphotonics.8b00151},
   url = {https://doi.org/10.1021/acsphotonics.8b00151},
   year = {2018},
   type = {Journal Article}
}

@article{Kojima2009,
author = {Kojima, Akihiro and Teshima, Kenjiro and Shirai, Yasuo and Miyasaka, Tsutomu},
title = {Organometal Halide Perovskites as Visible-Light Sensitizers for Photovoltaic Cells},
journal = {J. Am. Chem.},
volume = {131},
number = {17},
pages = {6050-6051},
year = {2009},
doi = {10.1021/ja809598r},

}

@article{Kim2025,
author = {Dasom Kim  and Sohail Dasgupta  and Xiaoxuan Ma  and Joong-Mok Park  and Hao-Tian Wei  and Xinwei Li  and Liang Luo  and Jacques Doumani  and Wanting Yang  and Di Cheng  and Richard H. J. Kim  and Henry O. Everitt  and Shojiro Kimura  and Hiroyuki Nojiri  and Jigang Wang  and Shixun Cao  and Motoaki Bamba  and Kaden R. A. Hazzard  and Junichiro Kono },
title = {Observation of the magnonic {D}icke superradiant phase transition},
journal = {Sci. Adv.},
volume = {11},
number = {14},
pages = {eadt1691},
year = {2025},
doi = {10.1126/sciadv.adt1691},
URL = {https://www.science.org/doi/abs/10.1126/sciadv.adt1691}}

@article{Scalari2012,
   author = {Scalari, G. and Maissen, C. and Turčinková, D. and Hagenmüller, D. and De Liberato, S. and Ciuti, C. and Reichl, C. and Schuh, D. and Wegscheider, W. and Beck, M. and Faist, J.},
   title = {Ultrastrong Coupling of the Cyclotron Transition of a 2{D} Electron Gas to a THz Metamaterial},
   journal = {Science},
   volume = {335},
   number = {6074},
   pages = {1323-1326},
   DOI = {doi:10.1126/science.1216022},
   url = {https://www.science.org/doi/abs/10.1126/science.1216022
https://www.science.org/doi/pdf/10.1126/science.1216022?download=true},
   year = {2012},
   type = {Journal Article}
}

@Article{Li2018,
  author   = {Xinwei Li and Motoaki Bamba and Ning Yuan and Qi Zhang and Yage Zhao and Maolin Xiang and Kai Xu and Zuanming Jin and Wei Ren and Guohong Ma and Shixun Cao and Dmitry Turchinovich and Junichiro Kono},
  journal  = {Science},
  title    = {Observation of {D}icke cooperativity in magnetic interactions},
  year     = {2018},
  number   = {6404},
  pages    = {794-797},
  volume   = {361},
  doi      = {10.1126/science.aat5162},
  url      = {https://www.science.org/doi/abs/10.1126/science.aat5162},
}

@ARTICLE{Kritzell2024-rw,
  title         = "Zeeman polaritons as a platform for probing {D}icke physics in
                   condensed matter",
  author        = "Kritzell, T Elijah and Doumani, Jacques and Asano, Tobias and
                   Yamada, Sota and Tay, Fuyang and Xu, Hongjing and Yan, Han
                   and Katayama, Ikufumi and Takeda, Jun and Nevidomskyy, Andriy
                   and Nojiri, Hiroyuki and Bamba, Motoaki and Baydin, Andrey
                   and Kono, Junichiro",
  journal       = "arXiv [quant-ph]",
  month         =  sep,
  year          =  2024,
  url           = "http://arxiv.org/abs/2409.17339",
  archivePrefix = "arXiv",
  primaryClass  = "quant-ph",
  eprint        = "2409.17339"
}

@article{hopfield_theory_1958,
	title = {Theory of the Contribution of Excitons to the Complex Dielectric Constant of Crystals},
	volume = {112},
	copyright = {http://link.aps.org/licenses/aps-default-license},
	issn = {0031-899X},
	url = {https://link.aps.org/doi/10.1103/PhysRev.112.1555},
	doi = {10.1103/PhysRev.112.1555},
	number = {5},
	urldate = {2024-10-09},
	journal = {Physical Review},
	author = {Hopfield, J. J.},
	month = dec,
	year = {1958},
	pages = {1555--1567},
}

@article{kim2025_nc,
  title = {Multimode Phonon-Polaritons in Lead-Halide Perovskites in the Ultrastrong Coupling Regime},
  author = {Kim, Dasom and Hou, Jin and Lee, Geon and Agrawal, Ayush and Kim, Sunghwan and Zhang, Hao and Bao, Di and Baydin, Andrey and Wu, Wenjing and Tay, Fuyang and Huang, Shengxi and Chia, Elbert E. M. and Kim, Dai-Sik and Seo, Minah and Mohite, Aditya D. and Hagenm{\"u}ller, David and Kono, Junichiro},
  year = 2025,
  month = sep,
  journal = {Nature Communications},
  volume = {16},
  number = {1},
  pages = {8658},
  issn = {2041-1723},
  doi = {10.1038/s41467-025-63810-7},
  url = {https://www.nature.com/articles/s41467-025-63810-7}
}

@article{Jarc2024,
    author = {Jarc, Giacomo and Mathengattil, Shahla Yasmin and Montanaro, Angela and Rigoni, Enrico Maria and Dal Zilio, Simone and Fausti, Daniele},
    title = {Multimode vibrational coupling across the insulator-to-metal transition in 1{T}-{TaS$_2$} in THz cavities},
    journal = {The Journal of Chemical Physics},
    volume = {161},
    number = {15},
    pages = {154711},
    year = {2024},
    month = {10},
    issn = {0021-9606},
    doi = {10.1063/5.0231533},
    url = {https://doi.org/10.1063/5.0231533},
}

@article{Jarc2023,
   author = {Jarc, Giacomo and Mathengattil, Shahla Yasmin and Montanaro, Angela and Giusti, Francesca and Rigoni, Enrico Maria and Sergo, Rudi and Fassioli, Francesca and Winnerl, Stephan and Dal Zilio, Simone and Mihailovic, Dragan and Prelovšek, Peter and Eckstein, Martin and Fausti, Daniele},
   title = {Cavity-mediated thermal control of metal-to-insulator transition in 1{T}-{TaS$_2$}},
   journal = {Nature},
   volume = {622},
   number = {7983},
   pages = {487-492},
   ISSN = {1476-4687},
   DOI = {10.1038/s41586-023-06596-2},
   url = {https://doi.org/10.1038/s41586-023-06596-2},
   year = {2023},
   type = {Journal Article}
}

@article{Paravicini2017,
  title = {Gate and magnetic field tunable ultrastrong coupling between a magnetoplasmon and the optical mode of an {LC} cavity},
  author = {Paravicini-Bagliani, Gian L. and Scalari, Giacomo and Valmorra, Federico and Keller, Janine and Maissen, Curdin and Beck, Mattias and Faist, J\'er\^ome},
  journal = {Phys. Rev. B},
  volume = {95},
  issue = {20},
  pages = {205304},
  numpages = {7},
  year = {2017},
  month = {May},
  publisher = {American Physical Society},
  doi = {10.1103/PhysRevB.95.205304},
  url = {https://link.aps.org/doi/10.1103/PhysRevB.95.205304}
}

\end{document}